\documentclass[aps,showpacs,prl,twocolumn,superscriptaddress]{revtex4}
\usepackage{epsfig}
\usepackage{amsfonts}
\usepackage{amsmath}
\usepackage[caption=false]{subfig}
\usepackage{float}
\usepackage{graphicx}
\usepackage[utf8]{inputenc}
\usepackage[colorlinks,bookmarks=false,citecolor=blue,linkcolor=blue,urlcolor=blue]{hyperref}

\newcommand{\ket}[1]{|#1\rangle}

\newcommand{\beq}{\begin{equation}}
\newcommand{\eeq}{\end{equation}}
\newcommand{\bea}{\begin{align}}
\newcommand{\eea}{\end{align}}
\newcommand{\nn}{\nonumber}

\newcommand{\ua}{\uparrow}
\newcommand{\da}{\downarrow}

\newcommand{\dagga}{{\phantom{\dagger}}}

\begin{document}
\title{Parity measurement in topological Josephson junctions }
\author{Fran\c{c}ois Cr\'epin}
\affiliation{Institute for Theoretical Physics and Astrophysics,
University of W\"urzburg, 97074 W\"urzburg, Germany}
\author{Bj\"orn Trauzettel}
\affiliation{Institute for Theoretical Physics and Astrophysics,
University of W\"urzburg, 97074 W\"urzburg, Germany}

\date{\today}

\begin{abstract}
We study the properties of a topological Josephson junction made of both edges of a 2D topological insulator. We show that, due to fermion parity pumping across the bulk, the global parity of the junction has a clear signature in the periodicity and critical value of the Josephson current. In particular, we find that the periodicity with the flux changes from $4\pi$ in a junction with an even number of quasi-particles to $2\pi$ in the odd sector. In the case of long junctions, we exhibit a rigorous mathematical connection between the spectrum of Andreev bound-states and the fermion parity anomaly, through bosonization. Additionally, we discuss the rather quantitative effects of Coulomb interactions on the Josephson current.
\end{abstract}


\maketitle

{\it Introduction.}  One-dimensional topological Josephson junctions are known to exhibit a $4\pi$-periodic Josephson current, a property that arises from a parity constraint on the number of quasi-particles in the junction~\cite{Kitaev01, Fu09b}. Depending on the precise system at hand, this effect can be related to the existence of Majorana (bound-)states, and hence provide a signature of Majorana fermions in condensed matter~\cite{Kitaev01, Fu09b}.  Two prominent proposals for the realization of 1D topological superconductors include spin-orbit wires in the presence of a magnetic field~\cite{Lutchyn10, Oreg10b}, as well as edge states of two-dimensional topological insulators (2D TI)~\cite{Fu09b}, both in proximity to an s-wave superconductor. Experimental evidence of a $4\pi$ Josephson current was recently reported in spin-orbit wires~\cite{Rokhinson12}.  A consequent body of work also describes possible realizations of qubits using MBS, as well as detailed quantum computation schemes~\cite{Nayak08, Alicea11}. A Majorana qubit consists of four Majorana fermions, which can be combined two by two into a state described by two fermion numbers only. By working in a sector of fixed parity, one can build a qubit using two out of the four available degenerate states.

\begin{figure}[h!] 
\centering
\includegraphics[width=4cm,clip]{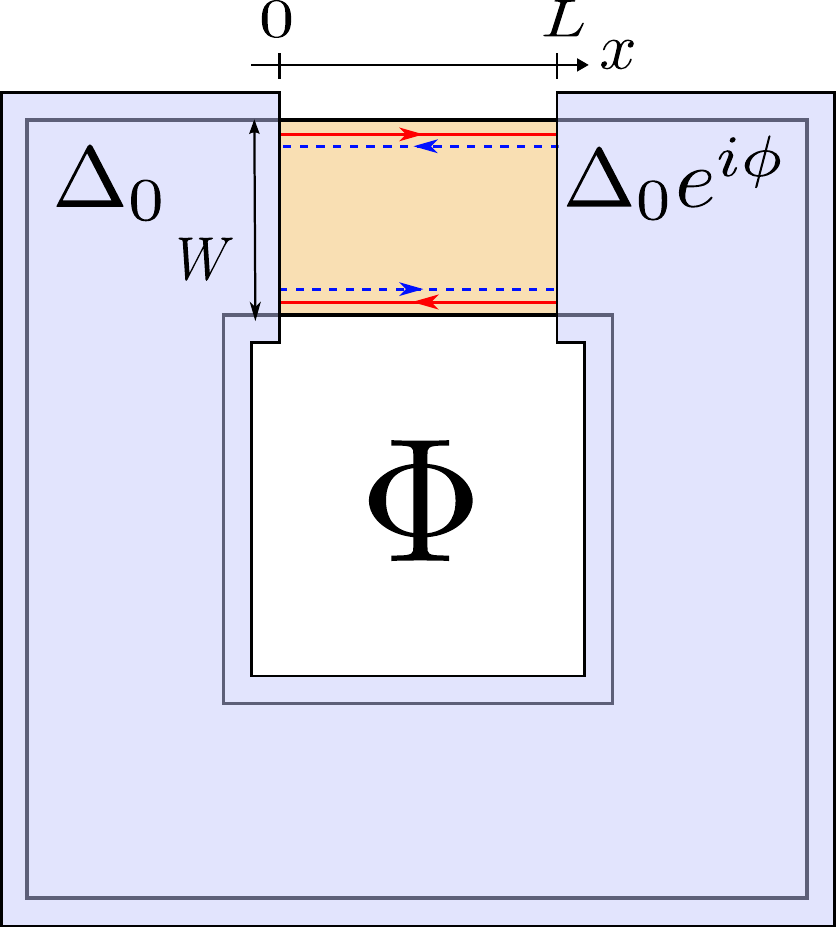}
\caption{ (Color online) Sketch of a topological Josephson junction. Both edges are contacted by the superconductor deposited on top of a 2D TI. In the simplest case where $S_z$ is conserved, solid red (resp. blue dashed) in the normal region indicate spin up (resp. spin down) electrons. A flux $\Phi$ through the hole induces a phase difference $\phi/(2\pi) = \Phi/\Phi_0$ across the Josephson junction with $\Phi_0 = h/2e$. }
\label{Fig:setup}
\end{figure}

In this Letter, we analyse a topological Josephson junction constructed on both edges of a ring-shaped 2D TI, as sketched in Fig.~\ref{Fig:setup}. 2D TIs have been realized in HgTe/CdTe quantum wells~\cite{Konig07}, as well as recently in InAs/GaSb systems~\cite{Knez11}, with normal and superconducting electrodes~\cite{Knez12}. In the normal phase, both edges are helical liquids and realize a state of perfect Andreev reflection at the NS interfaces. At first sight, this system is a peculiar realization of a spinful 1D Josephson junction~\cite{Fazio95, Maslov96, Affleck00} with the spin degrees of freedom split in two spatially separated regions. Although for large enough widths $W$ of the TI, tunneling between edges and inter-edge Coulomb interactions are exponentially suppressed, the two halves of the junction are connected by fermion parity pumping, as induced by the flux threaded through the ring~\cite{Fu06, Fu09b, Fu13}. 

We first analyse the parity constraints on the junction and propose to use properties of the Josephson current -- periodicity and critical value -- to measure the global parity of the junction. If we were to envision such a setup as a Majorana qubit, with two Majorana fermions at each edge, our method would in principle provide  a way to read out the parity of such a qubit~\footnote{We remark that a related setup was briefly discussed in Ref.~\onlinecite{Teo10b}, in the context of anyons, as a way to realize a so-called braidless operation of the type $\ket{00} \to \ket{11}$.}. We also present a detailed analysis of the long junction case through bosonization, discuss the effect of intra-edge interactions, and comment on the similarities and differences of this setup with respect to regular spinful Josephson junctions.

{\it Setup and parity measurement.} 
The system we have in mind is a Josephson junction consisting of a 2D topological insulator (TI) upon which an s-wave superconductor is deposited, as pictured in Fig.~\ref{Fig:setup}. In contrast with the original proposal of Ref.~\onlinecite{Fu09b}, and subsequent studies~\cite{Beenakker12}, we assume that a proximity gap $\Delta_0$ opens at both edges. 
In the normal region, a discrete spectrum of Andreev bound states (ABS) develops at each edge. Their energies depend on the phase difference $\phi$ across the junction, which can be induced by a magnetic flux.

For sufficiently large width of the TI, edge states on different edges do not interact and the Josephson current of the junction is given by the sum of two one-dimensional currents mediated by the upper and lower edges. Each current, taken separately, is a $4\pi$ periodic function of the phase difference, owing to the fermion parity anomaly~\cite{Fu09b} of the topological junction. The explanation of the anomaly relies on the following observation: while the edge Hamiltonian is $2\pi$ periodic in the phase difference $\phi$, its eigenstates are not. Due to particle-hole symmetry of the Bogoliubov-de Gennes Hamiltonian, each ABS at excitation energy $\varepsilon$ comes with a partner at $-\varepsilon$. These two states differ by their fermion parity. When the phase is advanced by $2\pi$, the system switches between them, and therefore changes parity: an extra quasi-particle has been added to the edge. In an ideal environment, that is, in the absence of quasi-particle poisoning, parity is conserved and the system can only return to its ground state by advancing again the phase by $2\pi$, thereby adding (or removing) a quasi-particle and restoring the parity. The origin of the extra quasi-particle lies in the topology of the 2D band insulator, and the existence of a second pair of edge states. In the geometry of Ref.~\onlinecite{Fu09b}, as $\phi$ is advance by $2\pi$,  a unit of fermion parity is pumped across the sample, from one edge to the other. In the present case of Fig.~\ref{Fig:setup}, this so-called  $\mathbb{Z}_2$ pumping~\cite{Fu09b, Fu06} effectively transfers a quasi-particle between ABS at the lower and upper edges. As a consequence, even though the fermion parity at each edge changes when $\phi$ advances by 2$\pi$, the global fermion parity of the junction is conserved.

 We write the Josephson current $J$ of the junction as
\beq
J_{\sigma,\sigma'} [\phi]= I_{\textrm{up},\sigma}[\phi] + I_{\textrm{down},\sigma'}[\phi]\;, \label{eq:current_1}
\eeq
where $\sigma = \pm$ (resp. $\sigma'=\pm$) indicates the fermion number parity ($+$ being for even and $-$ for odd) of the upper (resp. lower) edge. The total fermion parity of the junction is $\Sigma = \sigma \sigma'$. In Eq.~\eqref{eq:current_1} and in the following, we use the letter $J$ for the total current and the letter $I$ for the edge currents. A striking feature of the setup is that the periodicity of the Josephson current depends on the parity of the junction. For an even fermion number ($\Sigma = 1$), the current switches between $J_{+,+}$ and $J_{-,-}$ at $\phi = \pi \mod 2\pi$, and is thus $4\pi$-periodic in $\phi$. On the other hand, the odd current ($\Sigma = -1$) is $2\pi$-periodic. Indeed, it switches at $\phi = \pi \mod 2\pi$ between $J_{+,-}$ and $J_{-,+}$, which are equal functions, provided the upper and lower edges are identical, that is, $I_{\textrm{up},\sigma}[\phi] = I_{\textrm{down},\sigma}[\phi]$.

In addition, the parity of the junction is reflected in the value of the critical current $J_{\Sigma,c}$. Indeed, at a given edge, a change of parity reverses the direction of the current, that is,  $\textrm{sign}\;  I_{\textrm{up},+}[\phi] = - \textrm{sign} \; I_{\textrm{up},-}[\phi]$, and a similar relation for the lower edge. In the even sector, currents at both edges flow in the same direction, while in the odd sector they flow in opposite directions. As a consequence, the value of the critical current in the odd case, $J_{-,c}$ is reduced with respect to that in the even case, $J_{+,c}$. The exact ratio $J_{-,c}/J_{+,c}$ between even and odd critical currents depends on the length $L$ of the junction. In the short junction regime, $L \ll \xi$ with $\xi = v_F/\Delta_0$ the coherence length of the superconducting region, one has, at zero temperature, $ I_{\textrm{up},\pm}[\phi] =  I_{\textrm{down},\pm}[\phi] =  \pm I_c \sin (\phi/2)$ \cite{Fu09b}. Therefore, the even critical current is just $J_{+,c} = 2 I_c$, while the odd critical current vanishes, $J_{-,c} = 0$. Note that the odd current actually vanishes for all values of $\phi$ since $I_{\textrm{up},\sigma}[\phi] = - I_{\textrm{down},-\sigma}[\phi]$. The long junction regime, $L \gg \xi$, was studied in Ref.~\onlinecite{Beenakker12}. In that case one finds $ I_{\textrm{up},+}[\phi] =  I_{\textrm{down},+}[\phi] =  I'_c \phi/(2\pi)$ and  $ I_{\textrm{up},-}[\phi] =  I_{\textrm{down},-}[\phi] =  I'_c (\phi/(2\pi)- \textrm{sign} \; \phi)$. As a consequence, $J_{+,c} = 2 I'_c $ and $J_{-,c} =  I'_c $. In summary, the ratio of critical currents in the odd and even case interpolates between $J_{-,c}/J_{+,c} = 0$, for $L \ll \xi$ and $J_{-,c}/J_{+,c} = 1/2$, for $L \gg \xi$.

{\it Long junction regime and bosonization.}  The long junction regime is interesting for two reasons. First, it should be of experimental relevance, as argued in Ref.~\onlinecite{Beenakker12}. Second, it allows to make a clear mathematical connection between the spectrum of ABS and the fermion parity anomaly, through bosonization. At a given edge, helicity imposes perfect Andreev reflection and a non-degenerate spectrum of ABS. In the low-energy limit, $\varepsilon \ll \Delta_0$, the ABS spectrum is given by~\cite{Kulik70}
\beq
\varepsilon = \frac{ \pi v_F}{L}\left(n + \frac{1}{2} \pm \frac{\phi}{2\pi} \right)\; \qquad \textrm{with} \qquad n \in \mathbb{Z}\;. \label{eq:ABS_spectrum}
\vspace*{0.2cm}
\eeq
In the case of a non-helical, one-dimensional Josephson junction, this spectrum would be twice degenerate due to the spin 1/2 of the electrons. On a 2D topological insulator, the spin degree of freedom of the junction is splitted between the two edges, which are separated by the insulating bulk. As was first recognized in Ref.~\onlinecite{Maslov96}, the spectrum of Eq.~\eqref{eq:ABS_spectrum} is the one of 2 branches of counter-propagating fermions with a linear spectrum and twisted boundary conditions (TBC) on a segment of size $2L$. Indeed, at low-energy, one can write energy-independent boundary conditions for the fermion operators. At the upper edge, $\psi_{R,\ua}^\dagga(0) = i \psi_{L,\da}^\dagger(0)\;$ and 
$\psi_{R,\ua}^\dagga(L) = -i e^{i\phi}\psi_{L,\da}^\dagger(L) $ which implies 
\begin{subequations}
\begin{align}
&\psi_{R,\ua}^\dagga(x+2L,t) = - e^{-i\phi}\psi_{R,\ua}^\dagga(x,t)\;, \label{eq:twistedPBC1} \\
&\psi_{L,\da}^\dagga(x+2L,t) = - e^{-i\phi}\psi_{L,\da}^\dagga(x,t)\;, \label{eq:twistedPBC2} 
\end{align}
\end{subequations}
as well as a relation between right and left movers, 
\beq
\psi_{R,\ua}^\dagga(x,t) = -i \psi_{L,\da}^\dagger(-x,t)\;, \label{eq:chiral}
\eeq for $x \in [-L,L]$. One can then write an effective Hamiltonian for the normal region only with the appropriate boundary conditions. Starting from the Hamiltonian of the normal region, at the upper edge, we obtain
\beq
H_{\textrm{up}} = \frac{1}{2} \int_0^L dx \; \Psi^\dagger(x) \mathcal{H}_{\textrm{BdG}} \Psi(x)\;  \label{eq:H_1}
\eeq
with $\mathcal{H}_{\textrm{BdG}} = -i v_F \partial_x \sigma_z \tau_z $ the Bogoliubov-de Gennes Hamiltonian, and  $\Psi = ( \psi^\dagga_{R,\ua}, \psi^\dagga_{L,\da}, \psi^\dagger_{L,\da}, -\psi^\dagger_{R,\ua} )^T $. $\sigma_z$ (resp. $\tau_z$) is a Pauli matrix acting on the spin-1/2 (resp. particle-hole) degree of freedom. At low energies, using Eq.~\eqref{eq:chiral}, the Hamiltonian of Eq.~\eqref{eq:H_1} can be unfolded on the segment $[-L,L]$ as 
\beq
H_{\textrm{up}}= i \frac{v^\dagga_F}{2} \int_{-L}^L dx \; \left[-\psi_{R,\ua}^\dagger\partial_x \psi_{R,\ua}^\dagga +   \psi_{L,\da}^\dagger\partial_x \psi_{L,\da}^\dagga\right](x)\;, \label{eq:H_2}
\eeq
with the TBC of Eqs.~\eqref{eq:twistedPBC1} and \eqref{eq:twistedPBC2}. We then recover the spectrum of Eq.~\eqref{eq:ABS_spectrum}. The particle-hole symmetry can be made explicit upon Fourier transform and further use of Eq.~\eqref{eq:chiral}. Taking the superconducting gap to infinity, $\Delta_0 \rightarrow \infty$, we derive the following normal-ordered form of the Hamiltonian
\beq
H_{\textrm{up}} = \sum_{n \in \mathbb{Z}} \frac{v^\dagga_F\pi}{2L}\left(n + \frac{1}{2} - \frac{\phi}{2\pi} \right) : c^\dagger_{R,n} c^\dagga_{R,n} - c^\dagga_{R,n} c^\dagger_{R,n}:\;, \label{eq:H_3}
\eeq
where $\psi_{R,\ua}^\dagga(x,t) = \frac{1}{\sqrt{2L}}\sum_{n} e^{ik_nx}c_{R,n}\;, $ with the momentum $k_n$ quantized as $k_n = (\pi/L) (n +1/2-\phi/(2\pi))$. The spectrum of ABS reduces to that of a single band of fermionic quasi-particles with states labeled by an integer $n$. In this form, states with opposite energies obvioulsy differ in their fermion parity. Note that normal-ordering substracts an infinite ground-state energy coming from quasi-particle states below the Fermi energy, $E_F = 0$. This constant is independant of the phase and therefore does not affect the Josephson current.

 As the spectrum is unbounded, the Hamiltonian of Eq.~\eqref{eq:H_3} is invariant under the transformation $\phi \rightarrow \phi +2\pi$ and $n-1 \rightarrow n$. This transformation leaves the spectrum invariant but changes the parity of the ground-state. In this form, the model satisfies all the criteria -- a band of fermions with an unbounded, discrete spectrum -- for rigorous bosonization~\cite{VonDelft98}. The bosonization transformation reads as
\beq
\psi^\dagga_{R,\ua}(x,t) \equiv F_R \frac{1}{\sqrt{2\pi a}} e^{i \frac{\pi}{L} \left(\hat{\mathcal{N}}_R +\frac{1}{2} - \frac{\phi}{2\pi} \right)x} e^{-i \tilde{\phi}_R(x) }\;. \label{eq:bosonization_identity}
\eeq
Here, $\tilde{\phi}_R(x)$ is a chiral bosonic field, periodic on $[-L,L]$, with the following Fourier decomposition
\beq
\tilde{\phi}_R(x) = -\sum_{q>0} \sqrt{\frac{\pi}{L q}}  \left(e^{iqx} b^\dagga_{R,q} + e^{-iqx} b^\dagger_{R,q}\right)e^{-a q/2}\;, \label{eq:boson_field}
\eeq
where  $q = \frac{n\pi}{L}, \; n \in \mathbb{N}^*$. The bosonic operators $b^\dagger_{R,q}$ and $b_{R,q}$ create and annihilate particle-hole excitations on top of the Fermi sea. $\hat{\mathcal{N}}_R$ is an operator counting the number of fermions with respect to the Fermi sea~\footnote{The number of fermions is defined algebraically, and therefore counts holes with a negative sign.}. $F_R$ is the so-called Klein factor that lowers the fermion number by one. The factor $a$ is a regularization factor. Upon normal-ordering $e^{-i \tilde{\phi}_R(x)}$, it can be safely taken to be zero. In practice however, it can be retained and plays the role of a high-energy cutoff on the otherwise unbounded theory. Here, one should take $a \simeq \xi = v_F/\Delta_0$. Using Eqs.~\eqref{eq:bosonization_identity} and \eqref{eq:boson_field}, the Hamiltonian~\eqref{eq:H_3} takes the bosonized form~\cite{VonDelft98}
\beq
H_{\textrm{up}}= \frac{v^\dagga_F \pi}{2 L}   \left(  \hat{\mathcal{N}}_R - \frac{\phi}{2\pi} \right)^2 +  v^\dagga_F \sum_{q>0} q \; b^\dagger_{R,q} b^\dagga_{R,q}\;. \label{eq:H_boso}
\eeq
Obviously, the transformation $\phi \rightarrow \phi + 2\pi$, changes the topological number $\mathcal{N}_R $ by 1, a reflection of the fermion parity anomaly. As the Hamiltonian now naturally separates between a phase-dependent topological sector -- the number of quasi-particles in the junction-- and a non-topological sector -- the phase-independent particle-hole excitations -- the computation of the current at one edge, taking into account the parity constraints, is straightforward:
\beq
I_{\textrm{up},\pm}[\phi] = - \frac{2e}{\hbar} \frac{1}{\beta} \frac{\partial}{\partial \phi} \ln Z^\textrm{t}_{\textrm{up},\pm}[\phi]\;
\eeq
with 
\beq
Z^\textrm{t}_{\textrm{up},\pm}[\phi] = \sum_{\substack{\mathcal{N}_R \in \mathbb{Z}\\ \textrm{even/odd}}} e^{-\beta  \frac{\hbar \pi v^\dagga_F}{2L}  (\mathcal{N}_R- \phi/(2\pi))^2}\;, \label{eq:part_func}
\eeq
the topological part of the partition function. The parity constraint is simply reflected in the parity of the quasi-particle number $\mathcal{N}_R $. Interestingly, these partition functions can be expressed analytically in the form of Jacobi's elliptic theta functions~\footnote{Jacobi theta functions are defined as $\theta_3(z,q) =1+ 2 \sum_{n=0}^\infty q^{n^2}\cos[2nz] $ and $\theta_2(z,q) = 2 q^{1/4}\sum_{n=0}^\infty q^{n(n+1)}\cos[(2n+1)z]$. See, e.g, NIST Digital Library of Mathematical Functions, http://dlmf.nist.gov/20.2 }, and we find for the currents,
\begin{align}
&I_{\textrm{up},\pm}[\phi] = e \frac{ v^\dagga_F }{L}  \frac{\phi}{2\pi} - \frac{2 e}{\hbar \beta}\partial_\phi \ln  \theta_{3/2}\left[z(\phi),q\right]\;, \label{eq:current_2}
\end{align}
for $ |\phi| < \pi$. We have defined $z(\phi) = i \beta \frac{ \hbar \pi v^\dagga_F}{2L} \frac{\phi}{\pi}$ and $q = e^{-2\beta  \frac{ \hbar \pi v^\dagga_F}{L}}$. We have checked that our Eq.~\eqref{eq:current_2} and the formula of Beenakker {\it et al.}~\cite{Beenakker12} for the long junction Josephson current, although seemingly different, coincide in the limit $\Delta_0 \to \infty$~\footnote{In the supplemetary material to this Letter, we plotted Eq.~\eqref{eq:current_2}, and checked that it matches the corresponding results of Ref.~\onlinecite{Beenakker12}. We also outlined some steps of the bosonization calculation, including the treatment of interactions.  }. In particular, the odd current develops a discontinuity at $\phi=0$, as $\beta \to \infty$. Our bosonization approach is restricted to long junctions, nevertheless it has the conceptual advantage of putting the emphasis on the fermion parity anomaly through a simple counting of quasi-particles, and leads to a very simple formula for the current.



\begin{figure}[h!] 
\centering
\includegraphics[width=8cm,clip]{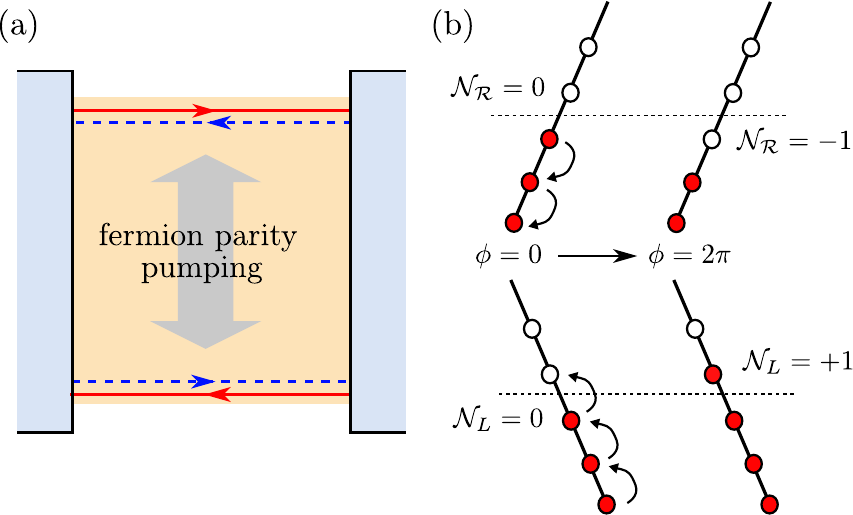}
\caption{(Color online) (a) Illustration of the fermion parity ($\mathbb{Z}_2$) pumping between the upper and lower junctions. (b) We show the linear spectra of Eqs.~\eqref{eq:H_3} and \eqref{eq:H_4}. At $\phi = 0$, we assume that all states with negative energy ($n < 0$) are filled and all states with positive energy ($ n \geq 0$) are empty, therefore $\mathcal{N}_R = \mathcal{N}_L = 0$. Following Eqs.~\eqref{eq:H_boso} and \eqref{eq:H_boso_2}, changing the flux by $2\pi$ transfers one quasi-particle from the upper edge to the lower edge. }
\label{Fig:spectrum}
\end{figure}

 Now, we include the lower-edge, which is most easily achieved by exchanging $\ua$ and $\da$ indexes in Eqs.~\eqref{eq:H_1} and \eqref{eq:H_2}. The effective low-energy Hamiltonian at the lower edge then reads
\beq
H_{\textrm{down}} = \sum_{n \in \mathbb{Z}} \frac{v^\dagga_F\pi}{2L}\left(n + \frac{1}{2} + \frac{\phi}{2\pi} \right) : c^\dagger_{L,n} c^\dagga_{L,n} - c^\dagga_{L,n} c^\dagger_{L,n}:\;, \label{eq:H_4}
\eeq
where $\psi_{L,\ua}^\dagga(x,t) = \frac{1}{\sqrt{2L}}\sum_{n} e^{-ik_nx}c_{L,n}\; $ with the momentum $k_n$  quantized as $k_n = (\pi/L) (n +1/2+\phi/(2\pi))$. In bosonized form,
\beq
H_{\textrm{down}}= \frac{v^\dagga_F \pi}{2 L}   \left(  \hat{\mathcal{N}}_L + \frac{\phi}{2\pi} \right)^2 +  v^\dagga_F \sum_{q>0} q \; b^\dagger_{L,q} b^\dagga_{L,q}\;. \label{eq:H_boso_2}
\eeq
Again, the transformation $\phi \rightarrow \phi + 2\pi$, changes the topological number $\mathcal{N}_L $ by 1. The combination of both Hamiltonians, \eqref{eq:H_boso} and \eqref{eq:H_boso_2}, illustrates the transfer of fermion parity from one edge to the other, as the phase is advanced by $2\pi$. A specific example is given in Fig.~\ref{Fig:spectrum}. Note that one could also write $H_{\textrm{down}}$ in terms of right movers only. This freedom of choice comes from Eq.~\eqref{eq:chiral}, and is due  to particle-hole symmetry. It illustrates the important fact that only fermion parity is well defined at the edges. The partition function of the lower edge, $Z^\textrm{t}_{\textrm{down},\pm}[\phi]$, is given by analogy with Eq.~\eqref{eq:part_func}. One can check, by making the change of variables $\mathcal{N}_L \rightarrow - \mathcal{N}_L $, that $I_{\textrm{down},\pm}[\phi] = I_{\textrm{up},\pm}[\phi]$. The partition function of the junction is simply $Z^\textrm{t}_{\sigma,\sigma'} =  Z^\textrm{t}_{\textrm{up},\sigma} \times Z^\textrm{t}_{\textrm{down},\sigma'}$ and the Josephson current is indeed the one of Eq.~\eqref{eq:current_1}.

{\it Effects of interactions.} Coulomb interactions at the edges can easily be included in the long junction regime on the basis of the bosonization method. Assuming for now that inter-edge interactions are made negligeable by screening,  the Hamiltonian at either edge conserves its simple low-energy form with only renormalized coefficients:
\begin{subequations}
\begin{align}
&H_{\textrm{up}}= \frac{v^\dagga_{N,u} \pi}{2 L}   \left(  \hat{\mathcal{N}}_R - \frac{\phi}{2\pi} \right)^2 +  v^\dagga_{s,u} \sum_{q>0} q \; B^\dagger_{R,q} B^\dagga_{R,q}\;, \label{eq:H_boso_r}\\
&H_{\textrm{down}}= \frac{v^\dagga_{N,d} \pi}{2 L}   \left(  \hat{\mathcal{N}}_L + \frac{\phi}{2\pi} \right)^2 +  v^\dagga_{s,d} \sum_{q>0} q \; B^\dagger_{L,q} B^\dagga_{L,q}\;, \label{eq:H_boso_r2}
\end{align}
\end{subequations}
where we have allowed for the possibility of different interaction strengths at each edge. The new bosonic operators $B$ and $B^\dagger$ are related to the original $b$ and $b^\dagger$ by a Bogoliubov transformation~\cite{Fabrizio95}. Quite generally, for short-range interactions, we have $v^\dagga_{N,\alpha} = v^\dagga_F - g^\dagga_{2,\alpha}/(2\pi)$, where $g^\dagga_{2,\alpha}$ is the original interaction between right and left movers at each edge. The Josephson current is simply obtained by replacing $v^\dagga_F$ by $v^\dagga_{N,\alpha}$ in Eq.~\eqref{eq:current_2}. The effects of repulsive interactions is therefore to reduce the Josephson current, as was first discovered in the spinful case~\cite{Maslov96}. One should note that our original prediction on the relation between parity and periodicity will not hold in the case $v^\dagga_{N,u} \neq v^\dagga_{N,d}$, since it would imply $I_{\textrm{down},\pm}[\phi] \neq I_{\textrm{up},\pm}[\phi]$. However, the ratio of critical currents in the odd and even cases is unaffected by interactions. Similar conclusions can be reached in the case of equal interactions but unequal lengths of the junctions.

Another interesting question raised by this particular setup is its relation to the well-known case of a 1D spinful Josephson junction~\cite{Maslov96, Affleck00, Fazio95, Fazio96}. One difficulty in that case is to incorporate both Andreev reflection and backscattering at the edge. The bosonization treatment of Ref.~\onlinecite{Maslov96} was built on the assumption of perfect Andreev reflection, which needed to be corrected to account for normal reflection. The case of helical liquids provide a natural realization of the assumption of perfect Andreev reflection. The present setup can be thought of as a way to split the ABS of the spinful liquid in two halves, spatially localized in two different regions, the edges, separated by the bulk gap. As we have seen, although the spin degeneracy is recovered, contrary to the single-edge junction of Ref.~\onlinecite{Fu09b}, the parity pumping across the topological insulator entails very different predictions for the Josephson current. Note that, in a situation where the fermion parity would not be conserved -- for instance if quasi-particle poisoning is important \cite{Budich12d, Rainis12} -- one simply recovers the Josephson current of the spinful liquid (a sawtooth with period $2\pi$), by letting the fermion number in Eq.~\eqref{eq:part_func} run over both even and odd integers. 

{\it Discussion}. In summary, we have studied a topological Josephson junction involving both edges of a 2D TI. Due to fermion parity pumping across the bulk, this system  differs greatly from the usual 1D spinful junction. We have put forward a possible non-destructive measurement of the global parity, that includes the effect of intra-edge interactions. As a closing remark, let us briefly elaborate on inter-edge interactions. So-called $g_{1, \parallel}$ processes~\cite{GiamarchiBook} do not change the local fermion number and can be easily incorporated in the bosonization treatment, leaving our general conclusions unchanged. On the other hand, upon inclusion of $g_{1, \perp}$ processes, the Hamiltonian no longer commutes with the fermion number, and a simple expression of the partition function as in Eq.~\eqref{eq:part_func} {\it a priori} no longer exists. Another interesting open question is the influence of disorder on our results \cite{Beenakker13}, in which case the two Josephson currents could be separately modified.

{\it Acknowledgments.} We thank Fabrizio Dolcini and Carlo Beenakker for illuminating discussions. Financial support by the DFG (German-Japanese research unit "Topotronics" and the priority program "Topological insulators") as well as the Helmholtz Foundation (VITI) is gratefully acknowledged.

\bibliography{Top_ins_wurzburg}

\onecolumngrid
\appendix

\section{Supplementary material}

In this supplementary material, we outline some steps about the various transformations leading from the Andreev bound-states (ABS) spectrum to the bosonized form of the junction Hamiltonian. Our approach is based on the work by Maslov {\it et al.} \cite{Maslov96} on spinful 1D Josephson junctions, yet we use somewhat different bosonization conventions that are specified below.

\section{A. Bosonization of the ABS spectrum}

For clarity, we now concentrate on the upper edge, where we have assumed that right movers have spin up and left movers spin down. In the junction, a right moving electron can be reflected as a left moving hole, or a left moving electron as a right moving hole. At a given energy $\varepsilon$ and phase difference $\phi$ across the junction, these processes give rise to two kinds of bound states with the constraint
\beq
2 \varepsilon L/v^\dagga_F \pm \phi - 2 \textrm{arccos} (\varepsilon/\Delta_0) = 2\pi n\;, \quad n \in \mathbb{Z}\;.
\eeq
In the low-energy regime, $ \varepsilon \ll \Delta_0$, the spectrum of ABS becomes linear,
\beq
\varepsilon_n = \frac{ \pi v_F}{L}\left(n + \frac{1}{2} \pm \frac{\phi}{2\pi} \right)\;,  \qquad n \in \mathbb{Z}\;, \label{eq:ABS_spectrum_app}
\vspace*{0.2cm}
\eeq
and one can write effective boundary conditions for the fermionic operators,
\begin{subequations}
\begin{align}
&\psi_{R,\ua}^\dagga(x+2L,t) = - e^{-i\phi}\psi_{R,\ua}^\dagga(x,t)\;, \label{eq:twistedPBC1_app} \\
&\psi_{L,\da}^\dagga(x+2L,t) = - e^{-i\phi}\psi_{L,\da}^\dagga(x,t)\;, \label{eq:twistedPBC2_app} 
\end{align}
\end{subequations}
as well as a relation between right and left movers, 
\beq
\psi_{R,\ua}^\dagga(x,t) = -i \psi_{L,\da}^\dagger(-x,t)\; \label{eq:chiral_app}
\eeq 
on an extended segment $[-L,L]$~\cite{Maslov96}. One can take advantage of equations \eqref{eq:twistedPBC1_app} and \eqref{eq:twistedPBC2_app} and unfold the Bogoliubov-de Gennes Hamiltonian of the normal region, on a fictitious ring of size $2L$,
\beq
H_{\textrm{up}}= i \frac{v^\dagga_F}{2} \int_{-L}^L dx \; \left[-\psi_{R,\ua}^\dagger\partial_x \psi_{R,\ua}^\dagga +   \psi_{L,\da}^\dagger\partial_x \psi_{L,\da}^\dagga\right](x)\; \label{eq:H_2_app}
\eeq
with twisted boundary conditions. Note that for $\phi=0$, $\psi_{R,\ua}$ and $\psi_{L,\da}$ satisfy anti-periodic boundary conditions. We introduce the following Fourier transforms for right and left movers,
\begin{subequations}
\begin{align}
&\psi_{R,\ua}^\dagga(x,t) = \frac{1}{\sqrt{2L}}\sum_{n} e^{i\frac{\pi}{L}\left(n + \frac{1}{2} + \frac{\phi}{2\pi} \right)x}c_{R,\ua,n}\;, \\
&\psi_{R,\ua}^\dagga(x,t) = \frac{1}{\sqrt{2L}}\sum_{n} e^{-i\frac{\pi}{L}\left(n + \frac{1}{2} - \frac{\phi}{2\pi} \right)x}c_{L,\da,n}\;.
\end{align}
\end{subequations}
The chiral constraint \eqref{eq:chiral_app} imposes $c^\dagga_{R,\ua,n} = -i  c^\dagger_{L,\da,-n-1}$. We then Fourier transform Eq.~\eqref{eq:H_2_app} into
\beq
H_{\textrm{up}} =  \frac{v_F}{2} \sum_{n \in \mathbb{Z}} \frac{\pi}{L}\left(n + \frac{1}{2} - \frac{\phi}{2\pi} \right)  c^\dagger_{R,\ua,n} c^\dagga_{R,\ua,n} +  \frac{\pi}{L}\left(n + \frac{1}{2} + \frac{\phi}{2\pi} \right)  c^\dagger_{L,\da,n} c^\dagga_{L,\da,n} \;. \label{eq:H_FT_1_app}
\eeq
In order to bosonize the Hamiltonian \eqref{eq:H_2_app}, one must let $\Delta_0 \rightarrow \infty$ and work with an unbounded spectrum. This approximation should hold at low enough temperatures, $T \ll \Delta_0$. Note that, as far as the computation of the Josephson current is concerned, this amounts to neglecting the continuum of quasi-particles above the gap. We now define the 0-(quasi)particle ground state as the state with filled fermionic levels below $E_F=0$, and substract its (infinite) energy value from the Hamiltonian. This procedure is equivalent to normal-ordering of Eqs.~\eqref{eq:H_2_app} and \eqref{eq:H_FT_1_app}~\cite{VonDelft98}. As stated in the main text, this (infinite) constant is independent of the phase $\phi$: normal-ordering does not affect the Josephson current, it is a mere redefinition of the energy scale. We can now use the exact bosonization relations~\cite{VonDelft98}, for right and left movers,
\begin{subequations}
\begin{align}
&\psi_{R,\ua}^\dagga(x,t) \equiv F_{R,\ua}\frac{1}{\sqrt{2\pi a}} e^{i \frac{\pi}{L} \left(\hat{\mathcal{N}}_{R,\ua} -\frac{1}{2} - \frac{\phi}{2\pi} \right)x} e^{-i \tilde{\phi}_{R,\ua}(x) }\;,\\
&\psi_{L,\da}^\dagga(x,t) \equiv F_{L,\da}\frac{1}{\sqrt{2\pi a}} e^{-i \frac{\pi}{L} \left(\hat{\mathcal{N}}_{L,\da} -\frac{1}{2} + \frac{\phi}{2\pi} \right)x} e^{+i \tilde{\phi}_{L,\da}(x) }\;.
\end{align}
\end{subequations}
Here, $\tilde{\phi}_{R,\ua}$ and $\tilde{\phi}_{L,\da}$ are two chiral bosonic fields, periodic on $[-L,L]$. $\hat{\mathcal{N}}_{R,\ua}$ and $\hat{\mathcal{N}}_{L,\da}$ are operators counting the number of quasi-particles with respect to the 0-(quasi)particle ground state. $ F_{R,\ua}$ and $ F_{L,\da}$ are the so-called Klein factors. The chiral relation \eqref{eq:chiral_app} implies
\begin{subequations}
\begin{align}
&F_{R,\ua} = -i F^\dagger_{L,\da}\;, \\
&\hat{\mathcal{N}}_{R,\ua} = -\hat{\mathcal{N}}_{L,\da} \;, \\
&\tilde{\phi}_{R,\ua}(x) = \tilde{\phi}_{L,\da}(-x)\;. \label{eq:bosonic_fields_app}
\end{align}
\label{eq:boso_constraints_app}
\end{subequations}
Then, the bosonized form of the Hamiltonian reads as
\begin{align}
H_{\textrm{up}} &= \frac{v^\dagga_F}{2}\frac{\pi }{2L}   \left(  \hat{\mathcal{N}}_{R,\ua} -   \frac{\phi}{2\pi} \right)^2 +\frac{v^\dagga_F}{2} \int_{-L}^L dx\; :\left( \partial_x \phi_{R,\ua}(x)\right)^ 2: + \nn \\
& + \frac{v^\dagga_F}{2}\frac{\pi }{2L}   \left(  \hat{\mathcal{N}}_{L,\da} +     \frac{\phi}{2\pi} \right)^2 +\frac{v^\dagga_F}{2} \int_{-L}^L dx\; :\left( \partial_x \phi_{L,\da}(x)\right)^ 2: \;,
\end{align}
which, using Eqs. \eqref{eq:boso_constraints_app}, can be expressed in terms of one species only (e.g. right movers with spin up) as
\beq
H_{\textrm{up}} =  \frac{v^\dagga_F \pi }{2L}   \left(  \hat{\mathcal{N}}_{R,\ua} -   \frac{\phi}{2\pi} \right)^2 +v^\dagga_F\int_{-L}^L dx\; :\left( \partial_x \phi_{R,\ua}(x)\right)^ 2:\;. \label{eq:H_boso_parity_app}
\eeq
This latter form emphasises the {\it particle-hole} symmetry of the ABS spectrum. Indeed, the Bogoliubov-de Gennes equations have particle-hole symmetry, that is, for each state with energy $\varepsilon$ there exists a state with energy $- \varepsilon$. This is apparent in Eq.~\eqref{eq:ABS_spectrum_app}, as $ \varepsilon_n = - \varepsilon_{-n-1}$, and can be made explicit in Eq. \eqref{eq:H_FT_1_app}, by making use of the chiral relation:
\beq
H_{\textrm{up}} = \sum_{n \in \mathbb{Z}} \epsilon_n(\phi)  : \left( 2 c^\dagger_{R,\ua, n} c^\dagga_{R,\ua, n} - 1 \right):\;, \label{eq:H_3_app}
\eeq
with $ \epsilon_n(\phi) = \frac{v^\dagga_F\pi}{2L}\left(n + \frac{1}{2} - \frac{\phi}{2\pi} \right)$. Thus, a single species of fermions is sufficient to describe the ABS spectrum, with the prescription that empty and filled levels have opposite energy. In the long junction regime, this formulation puts extra emphasis on the fermion parity. The Hamiltonian \eqref{eq:H_boso_parity_app} is the bosonized form of \eqref{eq:H_3_app}. Note that our choice to write the Hamiltonian in terms of right-movers with spin up only is arbitrary. Particle-hole symmetry of the BdG Hamiltonian also allows to work with left-movers with spin down only

 Actually, calling half of the ABS right movers and the other half left movers is also completely arbitrary. Indeed, they both consist of a superposition of particles and holes moving in opposite directions. Our choice nevertheless coincide with the direction of propagation of the supercurrent in the junction. Only the spin index is relevant since a given Andreev bound-state has a well defined spin projection in our problem.

\section{B. Intra-edge interactions}
We continue to focus on the upper-edge and introduce the following interaction Hamiltonian between right and left movers as,
\beq
H_{\textrm{int}} = g^\dagga_2 \int_0^L dx \; : \rho_{R,\ua}(x) \rho^\dagga_{L,\da}(x):\;,
\eeq
and extend it to the segment $[-L,L]$,
\beq
H_{\textrm{int}} = \frac{g^\dagga_2}{2} \int_0^L dx \; : \rho_{R,\ua}(x) \rho^\dagga_{L,\da}(x): + \frac{g^\dagga_2}{2} \int_{-L}^0 dx \; : \rho_{R,\ua}(-x) \rho^\dagga_{L,\da}(-x):\;.
\eeq
We then use $ \rho_{R,\ua}(x) = \mathcal{N}_{R,\ua}/(2L) + \partial_x \phi_{R,\ua}(x)/(2\pi) $ and  $ \rho_{L,\da}(x) = \mathcal{N}_{L,\da}/(2L) + \partial_x \phi_{L,\da}(x)/(2\pi) $, as well as Eq.~\eqref{eq:boso_constraints_app}, to find, after a few lines of algebra,
\beq
H_{\textrm{int}} = -  \frac{g^\dagga_2}{2} \frac{\hat{\mathcal{N}}_{R,\ua}^2}{2L} - \frac{g^\dagga_2}{2} \frac{1}{(2\pi)^2} \int_{-L}^L dx \; : \partial_x \phi_{R,\ua}(x) \partial_x \phi_{R,\ua}(-x) :\;.
\eeq
Let us point out that the interactions between quasi-particles have opposite sign with respect to the original interaction between fermions. A similar non-local interaction is obtained in a 1D wire with open boundary conditions~\cite{Fabrizio95}, however with the sign of interactions unchanged. This is essentially because in the open boundary case $ \mathcal{N}_{L,\da} = \mathcal{N}_{R,\ua}$ whereas in the present case $ \mathcal{N}_{L,\da} = -\mathcal{N}_{R,\ua}$. After a Bogoliubov transformation we arrive at the following bosonized Hamiltonian in the presence of interactions,
\beq
H_{\textrm{up}}= \frac{v^\dagga_{N} \pi}{2 L}   \left(  \hat{\mathcal{N}}_R - \frac{\phi}{2\pi} \right)^2 +  v^\dagga_{s} \sum_{q>0} q \; B^\dagger_{R,q} B^\dagga_{R,q}\;,
\eeq
with 
\beq
v^\dagga_N = v^\dagga_F (1 - g^\dagga_2/(2\pi v^\dagga_F ))\;,
\eeq
and
\beq
v^\dagga_s= v^\dagga_F \sqrt{1- g^{\dagga 2}_2/(2\pi v^\dagga_F )^2}\;.
\eeq
The Bogoliubov transformation that relates $ B^\dagger_{R,q}$ and $B^\dagga_{R,q}$ to $ b^\dagger_{R,q}$ and $b^\dagga_{R,q}$ is
\beq
b^\dagga_{R,q} = \cosh (\varphi) B^\dagga_{R,q} - \sinh (\varphi) B^\dagger_{R,q}
\eeq
with $\tanh (2\varphi) = g^\dagga_2/(\pi v^\dagga_F) $~\cite{Fabrizio95}.

\section{C. Plots of the Josephson current}

\begin{figure}
        \centering
        \subfloat[$E_T \beta = 1$]{\includegraphics[width=8cm,clip]{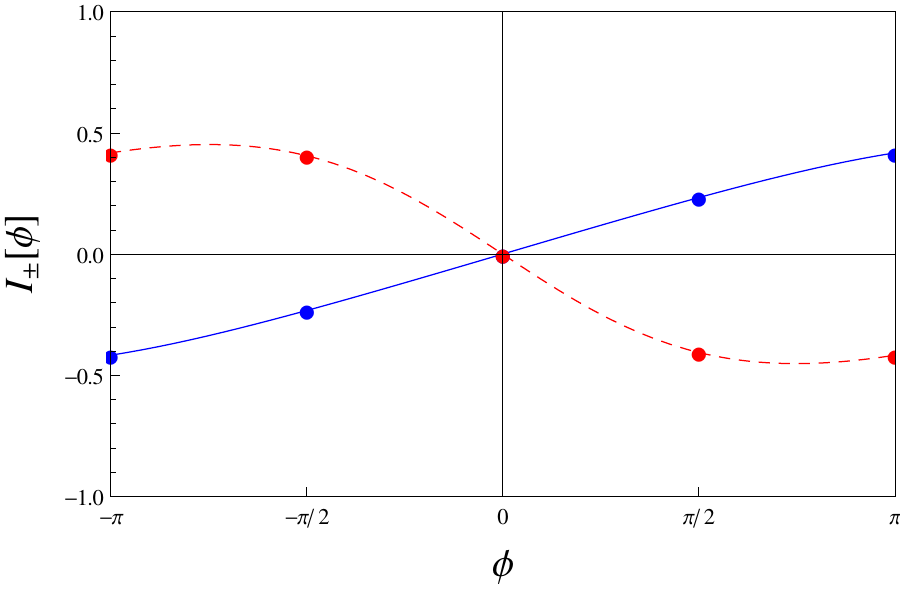}}
        \qquad
		\subfloat[$E_T \beta = 10$]{\includegraphics[width=8cm,clip]{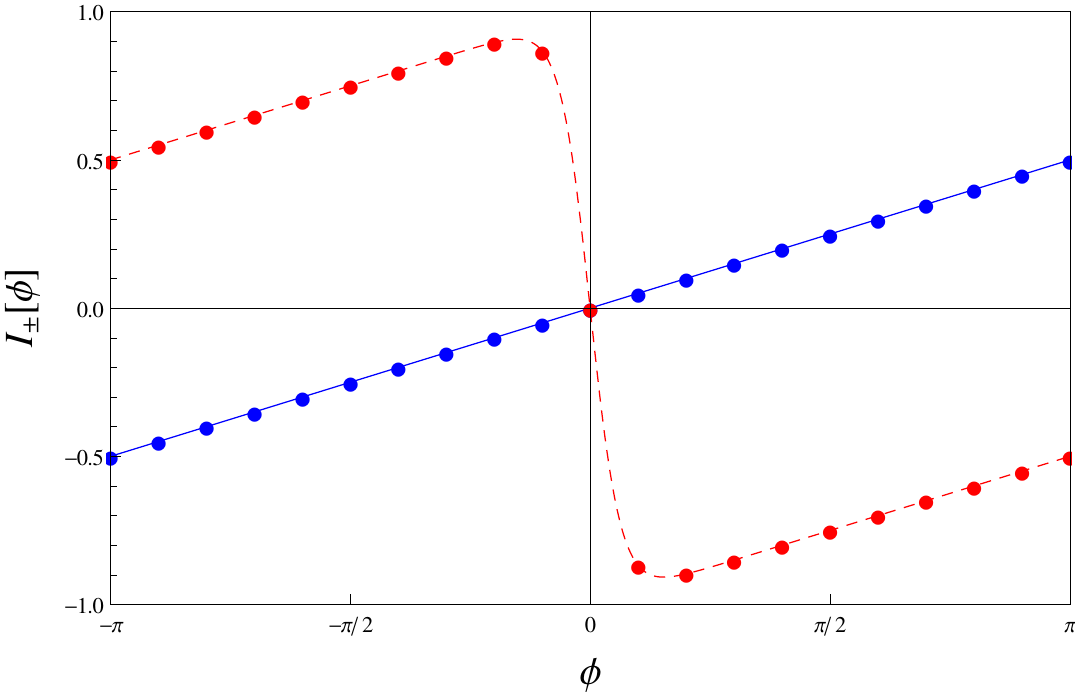}}
		\\
		\subfloat[$E_T \beta = 100$]{\includegraphics[width=8cm,clip]{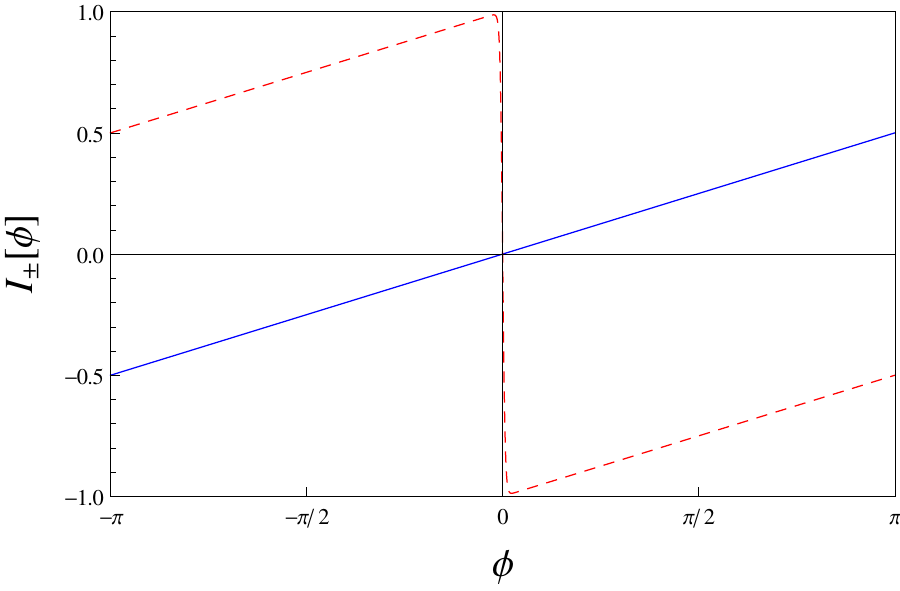}}
		\caption{Even (blue solid) and odd (red dashed) Josephson currents in the reduced zone, $-\pi<\phi<\pi$, in units of $\frac{e E_T}{\hbar}$, from Eq.~\eqref{eq:current_2_app}. As $E_T\beta \rightarrow \infty$, $I_-$ develops a discontinuity at $\phi = 0$, $I_+$ remains continuous. The filled circles correspond to the currents computed from the scattering method of Beenakker {\it et al.} (see main text). Evidently, the two methods match. }\label{fig:J}
\end{figure}

We now present a few plots of the Josephson current in the various situations described in the main text. First let us recall the expression for the current at one edge from Eq.~(13) of the main text,
\begin{align}
&I_{\textrm{up},\pm}[\phi] =  \frac{ e E_T }{\hbar}  \frac{\phi}{2\pi} - \frac{2 e}{\hbar \beta}\partial_\phi \ln  \theta_{3/2}\left[z(\phi),q\right]\;, \label{eq:current_2_app}
\end{align}
for $ |\phi| < \pi$. We have defined $z(\phi) = i \beta \frac{ \pi E_T}{2} \frac{\phi}{\pi}$ and $q = e^{-2\beta \pi E_T}$, and introduced the Thouless energy $E_T = \hbar v^\dagga_F/L$. The interacting result is obtained by replacing $v^\dagga_F$ by $v^\dagga_N$. In Fig.~\ref{fig:J} we compare our result to the results of Beenakker {\it et al.}~\cite{Beenakker12}, in the appropriate regime. Our formula reproduce their scattering formula in the limit $\Delta_0 \to \infty$ (see their Eq.~(18-20) and (A23)). The current outside of the reduced zone can be obtained by defining a piecewise function that switches between $I_{\textrm{up},+}[\phi]$ and $I_{\textrm{up},-}[\phi]$ at $\phi = \pi \mod 2\pi$. Note that Eq.~\eqref{eq:current_2_app} provides an alternative way as $I_{\textrm{up},+}[\phi + 2\pi] = I_{\textrm{up},-}[\phi]$. One can therefore plot  $I_{\textrm{up},+}[\phi]$ (or alternatively $I_{\textrm{up},-}[\phi]$) for all values of $\phi$ and still obtain the correct Josephson current. This is reminiscent of the short junction limit where $I_{\textrm{up},\pm}[\phi] = \pm \Delta_0/2 \sin(\phi/2)$  and obviously   $I_{\textrm{up},+}[\phi + 2\pi] = I_{\textrm{up},-}[\phi]$.  \\

\begin{figure}
        \centering
        \subfloat[$v_{N,u} = v_{N,d} = 1 $]{\includegraphics[width=8cm,clip]{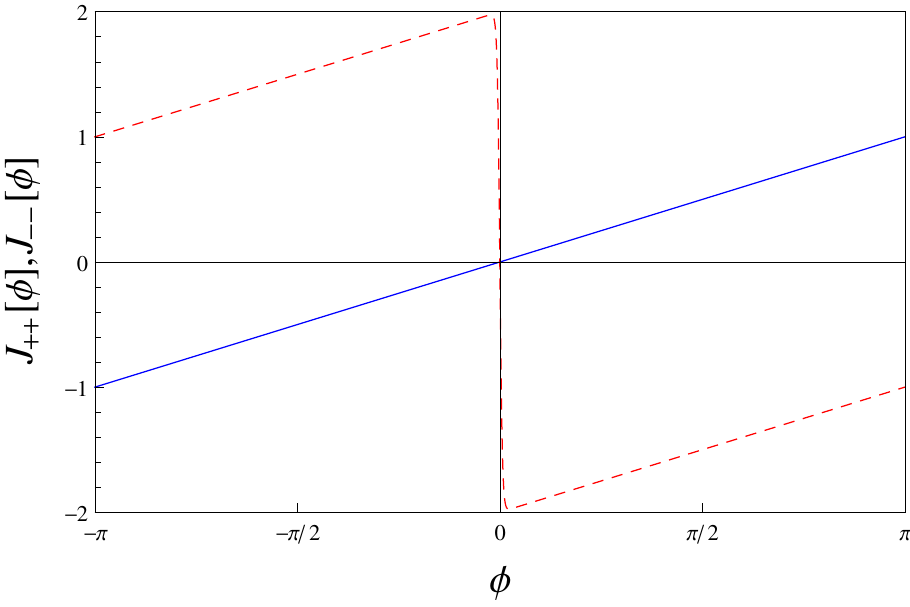}}
        \qquad
		\subfloat[$v_{N,u} = 0.8, \; v_{N,d} = 1 $]{\includegraphics[width=8cm,clip]{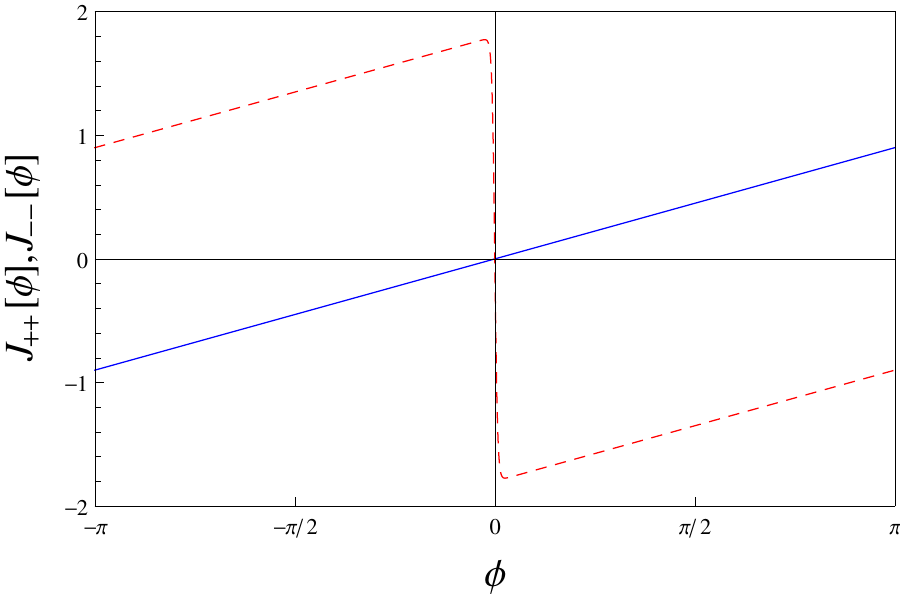}}
		\caption{Total Josephson current in the even sector, in units of $e E_T/\hbar$, with $E_T = \hbar v_F/L$, for equal and different values of $v_{N,u}$ and $v_{N,d}$. The temperature is fixed to $\beta = 100$. The two branches of the current, $J_{++}$ (blue solid) and $J_{--}$ (red dashed) are represented in the reduced zone $|\phi|<\pi$. The even current is $4\pi$ periodic in both cases.}\label{fig:J_total_even}
\end{figure}

\begin{figure}
        \centering
        \subfloat[$v_{N,u} = v_{N,d} = 1 $]{\includegraphics[width=8cm,clip]{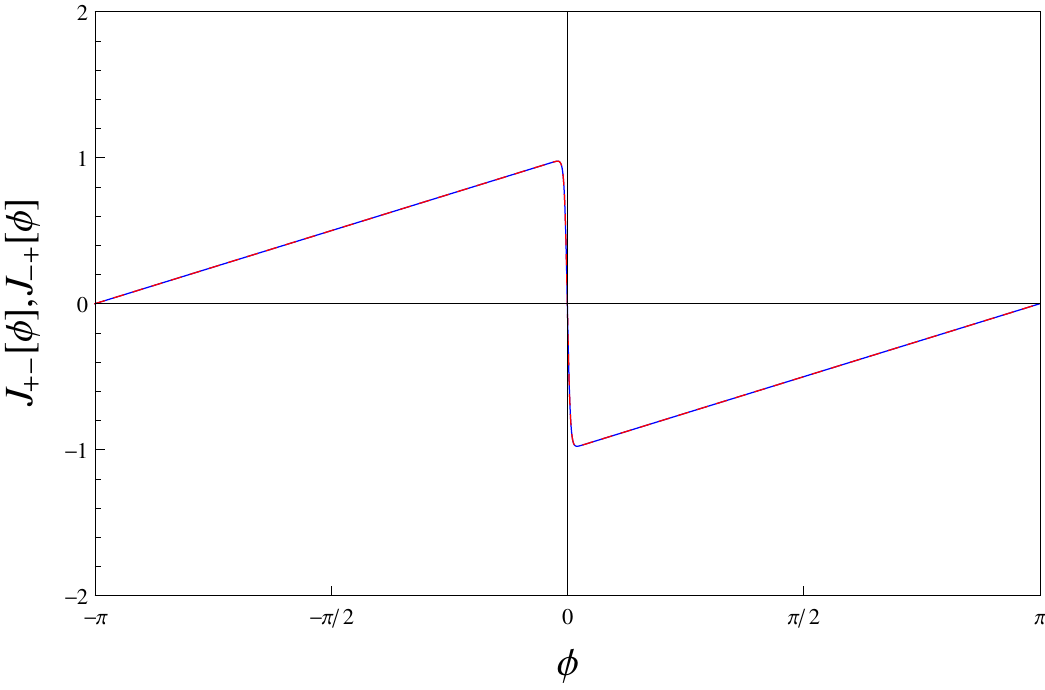}}
        \qquad
		\subfloat[$v_{N,u} = 0.8, \; v_{N,d} = 1 $]{\includegraphics[width=8cm,clip]{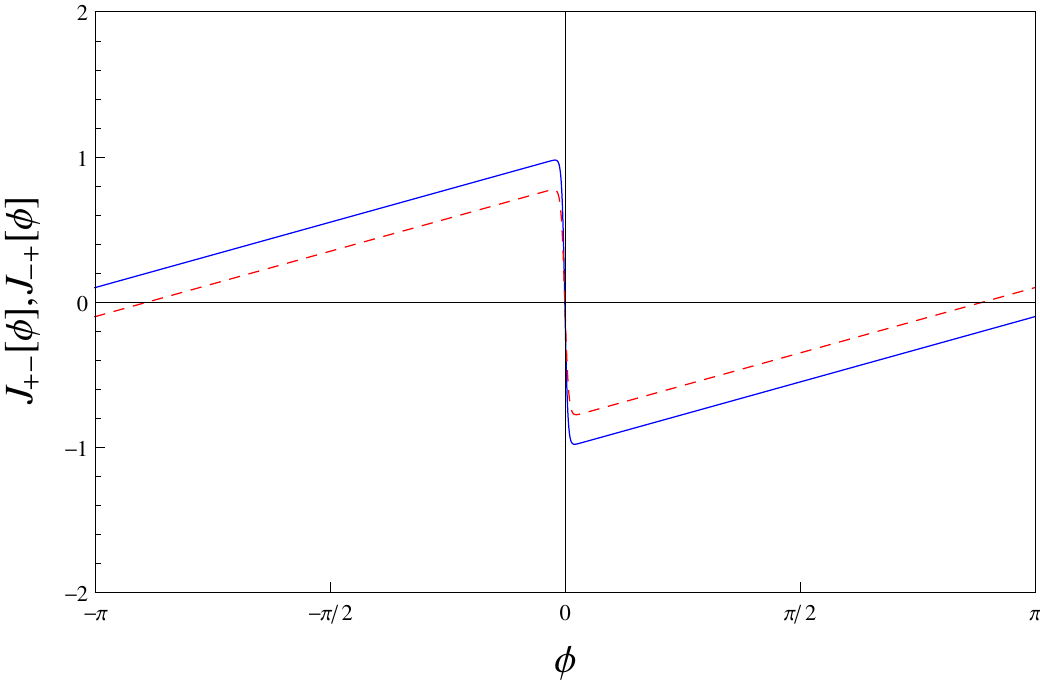}}
		 \caption{Total Josephson current in the odd sector, in units of $e E_T/\hbar$, with $E_T = \hbar v_F/L$, for equal and different values of $v_{N,u}$ and $v_{N,d}$. The temperature is fixed to $\beta = 100$. The two branches of the current, $J_{+-}$ (blue solid) and $J_{-+}$ (red dashed) are represented in the reduced zone $|\phi|<\pi$. For  $v_{N,u} = v_{N,d}$, $J_{+-} = J_{-+}$ and the odd current is $2\pi$ periodic, while for $v_{N,u} \neq v_{N,d}$, $J_{+-} \neq J_{-+}$ and the current is only $4\pi$ periodic.  }\label{fig:J_total_odd}
\end{figure}

		We next turn to the total current of the junction, that is, the sum of the upper and lower currents. In Figs.~\ref{fig:J_total_even} and \ref{fig:J_total_odd}, we have plotted the two branches of the current $J_{\Sigma}[\phi]$ in the reduced zone $|\phi|<\pi$ and in the long junction regime. At low temperatures, $E_T \beta \gg 1$, repulsive interactions, $v_{N,\alpha} < v_F$, tend to reduce the value of the critical current. An important point is that for  $v_{N,u} \neq v_{N,d}$, $J_{+-} \neq J_{-+}$ and the odd current is no longer $2\pi$ but instead $4\pi$ periodic. However the ratio of critical currents in even and odd sectors is unaffected.

\end{document}